\documentclass[preprint,12pt, nonatbib]{elsarticle}

\usepackage{amsfonts,amsmath,amssymb}
\usepackage{physics}
\usepackage{xcolor}
\usepackage{algorithm}
\usepackage{algorithmic}

\journal{Physics Letters A}

\begin{document}

\begin{frontmatter}



\title{Quantum optical model of an artificial neuron}

\author{Vivek Mehta}
\author{Utpal Roy}

\affiliation{organization={Department of Physics, Indian Institute of Technology Patna},
            addressline={Bihta},
            city={Patna},
            postcode={801103},
            state={Bihar},
            country={India}}

\begin{abstract}
Magnini \emph{et al.} [\emph{Mach. Learn.: Sci. Technol. 1 (2020) 045008}] recently introduced a qubit-based model of an artificial neuron, along with its applications. The design of its quantum circuit is pivotal for effective implementation. In this context, we present two quantum circuit synthesis algorithms tailored for the realisation of the quantum neuron. Comprehensive circuit simulations are conducted, and the resulting performance is assessed using the circuit cost metric. Additionally, we propose a quantum optical variant of the qubit-based quantum neuron, which offers a reduction in quantum resource requirements. To substantiate this, we introduce a quantum optical circuit synthesis algorithm and validate its efficacy through numerical simulations of prototype models.
\end{abstract}




\end{frontmatter}



\section{Introduction}
\label{sec1}
Deep neural networks have significantly advanced the field of artificial intelligence \cite{sarker2021deep}, powering applications such as large language models with billions of parameters \cite{shao2024survey}. Training these complex models necessitates substantial computational resources, prompting the need for efficient deployment strategies. One promising approach involves the development of specialized processing units capable of executing deep neural networks with reduced computational overhead. Quantum processing units (QPUs) offer such potential, leveraging quantum mechanical phenomena like superposition and entanglement to perform computations more efficiently than classical systems \cite{horodecki2009quantum,batin2024quantum, batin2024engineering, das2025quantum}. Recent theoretical advancements have introduced quantum neural network (QNN) algorithms that aim to minimize the computational demands of deep neural networks, thereby facilitating their deployment on quantum hardware \cite{rebentrost2018quantum, zhao2019building, killoran2019continuous}.


A neural network is a network of artificial neurons. This network was first conceptualized by observing the structure of the brain, constituting the networks of neuron cells. The objective function of an artificial neuron is to mimic the functionality of a neuron. Some quantum models of the artificial neuron are proposed with the intention for minimizing the computational resources of the classical artificial neuron \cite{schuld2015simulating,yan2020nonlinear, pechal2022direct}.  One such model is proposed by Mangini \textit{et al.} \cite{mangini2020quantum} for a continuously-valued input data with a prototype circuit, designed by the brute-force method. Consequently, a quantum implementation of feed-forward neural network can be designed using the scheme given by Tacchino. \textit{et al.} \cite{tacchino2020quantum}.

Therefore, it is important to come up with such quantum circuit synthesis algorithms for the implementation of the quantum neuron with arbitrary dimensional input data. Here, we explore two quantum synthesis algorithms and find out their circuit costs for implementation. Along with the detailed discussion on the implementation of the qubit-based model, we also provide a quantum optical model of this artificial neuron. The quantum optical circuit synthesis algorithm is also explicated for the quantum optical neuron. Advantages of the quantum optical model in comparison to the qubit-based model at the level of implementation are extracted by analysing the circuit cost metrics. Designing a quantum optical neuron model is further worthy for its feasibility to run over the integrated programmable scalable photonic hardware \cite{aghaee2025scaling}. The main reasons behind a photonic hardware to be extremely useful to implement various quantum machine learning algorithms are due to its advantages to operate at room temperature, requiring less energy during operation, and having longer coherence time \cite{mehta2024optical, schuld2019quantum, mehta2024quantum,anai2024continuous, mehta2024variable}.

The paper is outlined as follows. The next Sec. (\ref{sec2}) deals with a discussion on the qubit-based model of an artificial neuron as proposed by Mangini, \textit{et al.} \cite{mangini2020quantum}. We then discuss the qubit-based quantum circuit synthesis algorithms that generate circuits to implement a qubit-based quantum neuron. Numerical circuit simulation is also performed to explore the merits of these two qubit-based circuit synthesis algorithms. Section (\ref{sec3}) will comprise of a quantum optical model of the artificial neuron. The corresponding circuit synthesis algorithm and the numerical circuit simulation allow us to compute the circuit cost metric and a comparison with the qubit-based models before finally concluding in Sec. (\ref{sec4}).

\section{Qubit-based  Model of an Artificial Neuron}
\label{sec2}
\begin{figure*}
    \centering
    \includegraphics[width=.9\linewidth]{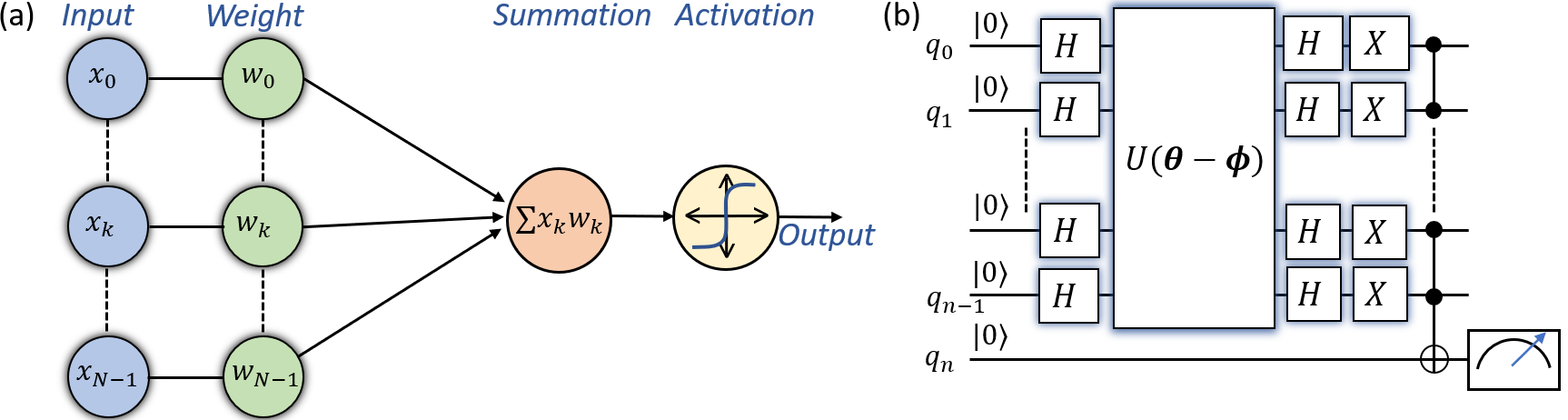 }
    \caption{Schematic of the implementation stages of an artificial neuron is shown in (a). An artificial neuron first computes the inner product between the input vector $\mathbf{x}$ and the weight $\mathbf{w}$, followed by an activation function to result an output. A qubit-based quantum neuron is shown in (b), which presents the stage-wise unitary evolution of the reference state embedding the input vector $\mathbf{x}$ and weight $\mathbf{w}$. The output of the quantum neuron is obtained upon measurement.}
    \label{fig1}
\end{figure*}

The salient steps towards implementing an artificial neuron is depicted in Fig. \ref{fig1}(a). A quantum model of an artificial neuron, as proposed by  Mangini \textit{et al.} \cite{mangini2020quantum}, also follows the same steps. For implementing these steps through a quantum circuit, we need to address a quantum model of an artificial neuron which can process continuous-valued information. A $N$-dimensional real space, $\mathbf{x}, \mathbf{w}\in \mathbb{R}^N$, comprises of the input and the weight vectors, which will be stored in the relative phases of the quantum wavefunction. Thus, each entry of these vectors is required to be scaled within $[0,\pi]$. A vector, $\mathbf{v}\in\mathbb{R}^N$, whose minimum and maximum values are respectively $a$ and $b$, can be scaled as
\begin{equation}
    \label{eq2.1}
    v_k\rightarrow \frac{v_k-a}{b-a}\times \pi,
\end{equation}
where the index $k$ ($0,\dots, N-1$) runs over all the components of the vector. Consequently, both the vectors ($\mathbf{x},\; \mathbf{w}$) transform into their \emph{rescaled forms} as $\boldsymbol{\theta}=(\theta_0, \dots, \theta_{n-1})$ and \emph{rescaled form} $\boldsymbol{\phi}=(\phi_0,\dots, \phi_{N-1})$, respectfully. These rescaled forms are used to define a \emph{new} set of input vector, $\mathbf{x}=(e^{i\theta_0}, \dots,e^{i\theta_{N-1}})$, and the  \emph{new} weight vector, $\mathbf{w}=(e^{i\phi_0}, \dots,e^{i\phi_{N-1}})$. When these vectors are stored in quantum states, they become
\begin{align}
    \ket{\Psi_{\mathbf{x}}}&=\frac{1}{\sqrt{2^n}}\sum_{k=0}^{2^n-1}e^{i\theta_k}\ket{k}, \text{ and}\label{eq2.2}\\
    \ket{\Psi_{\mathbf{w}}}&=\frac{1}{\sqrt{2^n}}\sum_{k=0}^{2^n-1}e^{i\phi_k}\ket{k},\label{eq2.3}
\end{align}
respectively, where $n=\lceil log_2N\rceil$ is the required number of qubits. In an artificial neuron, information is processed in two steps: an inner product is computed in the first place between the input and the weight and then, a nonlinear function is operated over the inner product. The inner product of the two states in Eq. (\ref{eq2.2}) and Eq. (\ref{eq2.3}) become
\begin{align}
    \langle\psi_{\mathbf{w}}|\psi_{\mathbf{x}}\rangle&=\frac{1}{2^n}\sum_{k=0}^{2^n-1}w^*_k x_{k} \notag \\
    &=\frac{1}{2^n}\sum_{k=0}^{2^n-1}e^{i(\theta_k-\phi_k)}.\label{eq2.4}
\end{align}
where `$*$' is used for complex conjugate. The corresponding quantum fidelity is written as
\begin{align}
   \left|\langle\psi_{\mathbf{w}}|\psi_{\mathbf{x}}\rangle\right|^2&=\left|\frac{1}{2^n}\sum_{k=0}^{2^n-1}e^{i(\theta_k-\phi_k)}\right|^2  \label{eq2.5}\\
   &=\frac{1}{2^n}+\frac{1}{2^{2n}}\sum_{j<k}cos\left((\theta_k-\phi_k)-(\theta_j-\phi_j)\right).\label{eq2.6}
\end{align}
Though it does not explicitly incorporate nonlinear or activation function, the representation of the phase-encoded quantum states is actually driven by an arbitrary nonlinear function and the quantum fidelity implicitly drives the activation function.

\subsection{Qubit-based Circuital Model of an Artificial Neuron}
\label{subsec2.1}
Quantum computation relies on unitary operations, where any unitary matrix can be decomposed into a sequence of one- and two-qubit quantum logic gates from a set of universal logic gates. A circuital quantum model of an artificial neuron is obtained by designing an appropriate sequence of unitary operations, followed by quantum measurements. The qubit-based implementation of an artificial neuron is shown in Fig. \ref{fig1} (b). The following unitary operations over the reference state $\ket{0}^{\otimes n}$ will generate the input quantum state
\begin{equation}
    \label{eq2.1.1}
    \ket{\Psi_{\mathbf{x}}}=U(\boldsymbol{\theta})H^{\otimes n}\ket{0}^{\otimes n},
\end{equation}
and the weight quantum state is found in a similar manner:
\begin{equation}
\label{eq2.1.2}
    \ket{\Psi_{\mathbf{w}}}=U(\boldsymbol{\phi})H^{\otimes n}\ket{0}^{\otimes n}.
\end{equation}
When the Hadamard gates $H^{\otimes n}$ operate over the state $\ket{0}^{\otimes n}$, it creates a uniform superposition:
\begin{equation}
    \label{eq2.1.3}
    H^{\otimes n}\ket{0}^{\otimes n}\rightarrow \frac{1}{\sqrt{2^n}}\sum_{k=0}^{2^n-1}\ket{k}.
\end{equation}
Subsequently, the unitary transformation, $U(\boldsymbol{\beta})$, with arbitrary angle vector $\boldsymbol{\beta}$, embeds the relative phase into the uniformly superposed quantum state.
\begin{equation}
    \label{eq2.1.4}
    U(\boldsymbol{\beta})\frac{1}{\sqrt{2^n}}\sum_{k=0}^{2^n-1}\ket{k}\rightarrow \frac{1}{\sqrt{2^n}}\sum_{k=0}^{2^n-1}e^{i\beta_k}\ket{k},
\end{equation}
which is referred to as a diagonal operator. Hence, the inner product as shown in Eq. (\ref{eq2.4}) can be written using Eqs. (\ref{eq2.1.1}) and (\ref{eq2.1.2}) as
\begin{equation}
    \label{eq2.1.5}
    \langle\psi_{\mathbf{w}}|\psi_{\mathbf{x}}\rangle=\text{}^{n\otimes }\langle 0|H^{\otimes n}U^\dagger(\boldsymbol{\phi})U(\boldsymbol{\theta})H^{\otimes n}\ket{0}^{\otimes n}.
\end{equation}
Here, the reference state $\ket{0}^{\otimes n}$ evolves with unitary transformation $H^{\otimes n}U^\dagger(\boldsymbol{\phi})$\\$U(\boldsymbol{\theta})H^{\otimes n}$ in proper sequence. It is worth combining both the diagonal operators, $U^\dagger(\boldsymbol{\phi})\text{ and }U(\boldsymbol{\theta})$, and get a resultant diagonal unitary operator $U(\boldsymbol{\theta}-\boldsymbol{\phi})$. This combination will help us to minimize the circuit complexity of the quantum circuit. Finally, the output state after the unitary evolution is projected on the zero computational basis using the projection operator $\ket{0}^{\otimes n \text{ }n\otimes}\langle 0|$. This projection operator can be implemented using the $n$ single-qubit Pauli-X (denotes as $X$) gates with one multiqubit $n$-controlled-NOT gate using an ancilla qubit, namely $q_n$ in Fig. \ref{fig1} (b).

\subsection{Quantum Circuit Synthesis for Diagonal Unitary Operators}
\label{subsec2.2}
The designing of the quantum circuit to implement the diagonal operator is crucial for the quantum circuital model of the quantum neuron. We are going to discuss \emph{two} qubit-based quantum circuit synthesis algorithms that generate those quantum circuits for realizing the diagonal operator. Consider a general diagonal operator in matrix form
\begin{equation}
    \label{eq2.2.1}
    U(\boldsymbol{\beta})=\begin{pmatrix}
        e^{-i\beta_0}&0&\dots&0\\
        0&e^{-i\beta_1}&\dots&0\\
        \vdots&\vdots&\ddots&\vdots\\
        0&0&\dots&e^{-i\beta_{N-1}}
    \end{pmatrix}_{N\times N}
\end{equation}
for $N=2^n$. An angle vector $\boldsymbol{\beta}=(\beta_0, \beta_1,\dots, \beta_{N-1})$ is related to another angle vector $\boldsymbol{\alpha}=(\alpha_0,\alpha_1,\dots,\alpha_{N-1})$ through an orthogonal and unitary matrix $M$ of $N\times N$ dimensions:
\begin{equation}
    \label{eq2.2.2}
    \boldsymbol{\beta}=M\boldsymbol{\alpha}.
\end{equation}
The entries $\alpha_k$ are actual rotation parameters corresponding to the logic gates in the required network that implements the unitary matrix $U(\boldsymbol{\beta})$. The two different structures of the matrix $M$ are the basis for two different quantum circuit synthesis algorithms. The following quantum circuit synthesis algorithm generates the quantum circuit for $U(\boldsymbol{\beta}$), where the matrix $M$ is defined and we construct the actual quantum circuit, comprising of \emph{elementary gate set} $\{R_z(\omega), Cnot\}$. The Pauli-Z rotation $R_z(\omega)$ for an arbitrary angle $\omega$ is written as
\begin{equation}
    \label{eq2.2.3}
    R_z(\omega)=\begin{pmatrix}
        e^{-\frac{i\omega}{2}}&0\\
        0& e^{\frac{i\omega}{2}}
    \end{pmatrix}
\end{equation}
and $Cnot$ is a two-qubit controlled Pauli-X gate.
\subsubsection{Algorithm I:}
Let us begin with structuring the matrix $M$ for Algorithm I:
\begin{equation}
    \label{eq2.2.1.1}
    M_{st}=\frac{(-1)^{b_s.g_t}}{2^n},
\end{equation}
where $b_s$ and $g_t$ are both the standard binary and Gray code representations of the integer $s,t=0,\dots, 2^n-1$, respectively and $b_s.g_t$ is the bitwise inner product between their binary representations.

Then, for discussing the structure of the quantum circuit, a quantum register of $(n+1)$ qubits, $\ket{0}^{\otimes n+1}$, are considered, where the last qubit is an ancilla qubit. The circuit is an alternative sequence of $\{R_z(-2\alpha_k)\}$ and $Cnot$ gates, where Pauli-Z gates act only on the ancilla qubit. The ancilla qubit is also the target qubit for all the control operations, nevertheless the controlled qubits are different. To determine the control qubits for all the $Cnot$ gates, we identify the control qubit of the $t$-th $Cnot$ gate, which is expressed by the position of the reflected Gray code between $g_{t-1}$ and $g_t$. We end up with a circuit for $U(\boldsymbol{\beta)}$ for method I, where the first gate in the circuit is $R_z(-2\alpha_0)$, acting on the ancilla qubit, followed by a $Cnot$ gate and further iterations.

\subsubsection{Algorithm II:}

The structure of the matrix $M$ for algorithm II (Schuch and Siewert \cite{schuch2003programmable}) is discussed. A recent discussion on the optimized circuital implementation of this algorithm by Zhang \textit{et al.} \cite{zhang2024depth} will be utilized. The matrix $M$ is an unnormalized Hadamard matrix of dimension $2^n\times 2^n$ whose elements are given by
\begin{equation}
    \label{eq2.2.2.1}
    M_{st}=\frac{(-1)^{b_s.b_t}}{2^{n-1}},
\end{equation}
where $b_s \text{ and } b_t$ are the standard binary representation of the integer $s,t=0,\dots, 2^n-1$, respectively, and $b_s.b_t$ is the bitwise inner product between their binary representations.

\begin{figure*}
    \centering
    \includegraphics[width=0.85\linewidth]{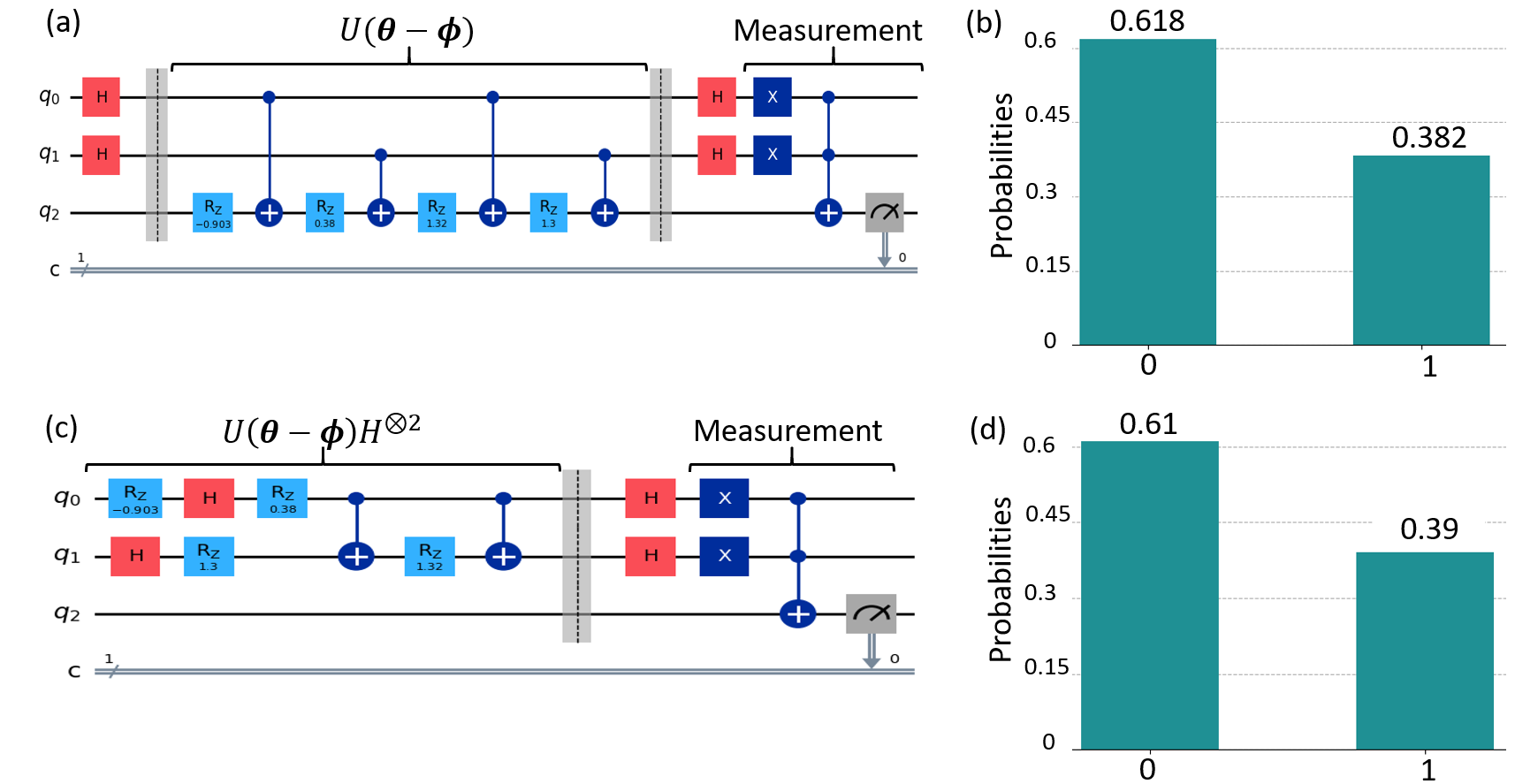}
    \caption{Figures (a) and (c) are quantum circuits that implement the qubit-based quantum model for four-dimensional input data and their diagonal operations, $U(\boldsymbol{\theta}-\boldsymbol{\phi})$, following algorithms I and II, respectively. These circuits are numerically simulated and the corresponding measurement outcomes are depicted in the histogram plots, (b) and (d). }
    \label{fig2}
\end{figure*}

For finding the structure of the quantum circuit, synthesized through algorithm II, we initialize a quantum register of $n$-qubit, $\ket{0}^{\otimes n}$. We apply $R_z(-\alpha_0)$ on one of the qubits with a global phase $\alpha_0$. Designing an artificial neuron needs a uniform superposition state in the computational basis, $\{0,1\}^n$. Here, the phase rotations will exclude $\alpha_0$. The indices $q$ of these phases are represented in the binary codes $\{q_{n-1}\dots q_1q_0\}$, where $q_{n-1}$ and $q_0$ are the most and least significant bits, respectively. Here, $q$ is selected as an arbitrary index. Let's start implementing the phase rotation with $\alpha_1$, that has binary bit representation $\alpha_{0\dots 1}$. We observe that the last bit is 1 and all other bits are 0's. As $q_{n-1}\dots q_1q_0$ is to represent the $n$-qubit quantum register, we apply a single $R_z(-\alpha_{0\dots 1})$ gate on the $q_0$ qubit without any associated $Cnot$ gates. To incorporate the successive rotations, $\alpha_{q_{n-1}\dots q_0}$, the following procedure is explicated. We need to see which leftmost bit has value 1 and then, we apply $R_z(-\alpha_{q_{n-1}\dots q_0})$ rotation gate on the qubit whose index matches with the index of the bit. Secondly, this $R_z(\alpha_{q_{n-1}\dots q_0})$ rotation gate may be associated with the $Cnot$ gates, depending on which bits also have value 1 other than the leftmost bit. If we get some bits of such values, the matching indices of qubits with these bits are the control qubits of the $Cnot$ gates, while their target qubits are the same $R_z(-\alpha_{q_{n-1}\dots q_0})$ rotation gates as were chosen. Notice that these $Cnot$ gates are applied in the same sequence on both sides of $R_z(\alpha_{q_{n-1}\dots q_0})$.

\subsection{Numerical Simulation of Quantum Circuits and their Circuit Costs for an Artificial Neuron in the Qubit-based Model}

Before simulating the quantum circuits of the artificial neuron, we will discuss about the circuit cost metrics. It is also known that a quantum circuit may be considered as a directed acyclic graph just like a classical Boolean circuit. In theoretical computer science, the circuit complexity is estimated by the \emph{circuit size} and the \emph{circuit depth}. The circuit size of a circuit signifies the number of constituent logic gates. On the other hand, circuit depth is defined as the longest path between an input and the output of a circuit.  Along with these metrics, the \emph{circuit width} is another representation of the cost metric, defined exclusively for quantum circuits by its number of qubits. However, the concept of circuit width is incompatible with the quantum optical circuit, but can be redefined for such cases \cite{arrazola2014quantum}. $log_2[dim(\mathcal{H})]$ represents the smallest number of qubits when the Hilbert space is associated with a quantum optical system.

With this, we can discuss the corresponding quantum circuit synthesis algorithms. Qiskit is a very well-known Python-based quantum computation platform, that is used to numerically simulate the qubit-based artificial model \cite{javadi2024quantum}. We consider the rescaled form of the input, $\mathbf{x}$ is $[\pi/6, \pi/3, \pi/2, \pi/5]$, for the weight $[\pi/2, \pi/8, 0, 0]$. Figures \ref{fig2} (a) and \ref{fig2} (c) depict two quantum circuits that are synthesized via algorithms I and II, respectively, which are numerically simulated with Qiskit. As a result, we get computational basis probabilities $p(0)$ and $p(1)$ which are shown in Figs. \ref{fig2} (b) and \ref{fig2} (d). The observed values of $p(1)$ in Figs. \ref{fig2} (b) and \ref{fig2} (d) are outputs of the quantum neuron model for the given input and weight vectors. The following points will help us to measure the merits of both the algorithms on the basis of circuit cost metrics. All these cost metrics are expressed in terms of the dimension of the input or the weight vector $N$.
\begin{enumerate}
    \item[(a)] \textbf{Circuit size}: The circuit sizes of those circuits, which are synthesized from the algorithms I and II are  $(3\lceil log_2 N\rceil+2^{(\lceil log_2 N\rceil+1)})$ and  $(3\lceil log_2 N\rceil+2^{(\lceil log_2 N\rceil+1)}-2)$, respectively, without including the multi-qubit controlled gates. However, the difference between the circuit sizes of both the algorithms I and II is only 2.
    \item[(b)] \textbf{Circuit depth}: The circuit depths of those circuits, which are synthesized from the algorithms I and II are $(2^{(\lceil log_2 N\rceil+1)}+3)$ and  $(2^{(\lceil log_2 N\rceil)}+3)$, respectively, without including the multi-qubit controlled gates. The circuit depth obtained from algorithm I is doubled of that from algorithm II.
    \item[(c)] \textbf{Circuit width}: Both the synthesis algorithms require the same number of qubits $ (\lceil log_2 N\rceil +1)$.
\end{enumerate}
Notice that, the multi-controlled gates are not required to implement a qubit-based quantum neuron. However, we need to measure all the $n$-qubits, which will increase the quantum resources and a quantum optical model becomes worth to try upon.

\section{Quantum Optical Model of an Artificial Neuron}
\label{sec3}
The quantum optical model relies on an integrated programmable quantum optical architecture, which represents the quantum optical states of the input and the weight:
\begin{align}
    \ket{\psi_{\mathbf{x}}}&=\frac{1}{\sqrt{N}}\sum_{k=0}^{N-1}e^{i\theta_k}\ket{1}_k\label{eq3.1}\\
    \ket{\psi_{\mathbf{w}}}&=\frac{1}{\sqrt{N}}\sum_{k=0}^{N-1}e^{i\phi_k}\ket{1}_k , \label{eq3.2}
\end{align}
where $\ket{1}_k$ is a single photon state in the $k$th \emph{spatial} quantum mode (in short `qmode'). Corresponding to each qmode, we can define the bosonic operators, $\hat{a}$ and $\hat{a}^\dagger$. These non-Hermitian operators follow the commutation relation; $[\hat{a}_k,\hat{a}^\dagger_{k'}]=\delta_{kk'}$, where $\delta$ is Kronecker delta.
It is worth drawing a parallel between the qubit-based and the optical-based Hilbert spaces for storing information. Both are discrete spaces, where the qubit-based space is described by the computational bases $\ket{k}$, whereas the optical-based space is described by the computational bases $\ket{1}_k$. When both the Hilbert spaces have the same dimension, $\ket{1}_k$ becomes equivalent to $\ket{k}$ for instance, a qubit $a\ket{1}_0+b\ket{1}_1$ is called a dual rail representation of the qubit $a\ket{0}+b\ket{1}$ \cite{nielsen2010quantum}. Hence, $\ket{1}_0$ is called $\ket{0}$ computational basis.

\begin{figure}
    \centering
    \includegraphics[width=.99\linewidth]{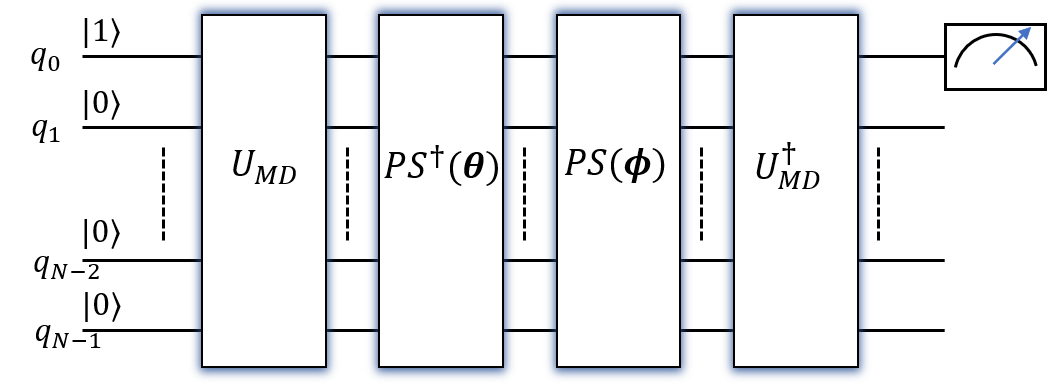}
    \caption{A quantum optical neuron, where the unitary evolution of the reference state, $\ket{1}_0$, embeds the classical information ($\mathbf{x}$ and $\mathbf{w}$). After the unitary evolution, the probability of getting no photon at the first mode is the output of the quantum neuron.}
    \label{fig3}
\end{figure}

Figure \ref{fig3} shows that the unitary operations evolve the reference state into the desirable quantum states, where the unitarity is through quantum optical networks. The following quantum optical operations over the reference state $\ket{1}_0$ will create the input quantum state
\begin{equation}
    \label{eq3.3}
    \ket{\psi_{\mathbf{x}}}=PS^\dagger(\boldsymbol{\theta)}U_{MD}(\{\eta_t\})\ket{1}_0
\end{equation}
and in the same manner, the weight is created as
\begin{equation}
    \label{eq3.4}
    \ket{\psi_{\mathbf{w}}}=PS^\dagger(\boldsymbol{\phi)}U_{MD}(\{\eta_t\})\ket{1}_0.
\end{equation}
Here, $U_{MD}(\{\eta_t\})$ is an optical multiport device (MD), which is made up of a number of beam splitters of different transmissivity coefficient $\eta_t$ with $t$ as an index. $PS(\boldsymbol{\beta}):=\bigotimes_{k=0}^{N-1}PS(\beta_k)$ is a tensor product of local phase shifters on overall spatial qmodes. We find the inner product between the states in Eqs. (\ref{eq3.1}) and (\ref{eq3.2}) using the optical operations in Eqs. (\ref{eq3.3}) and (\ref{eq3.4}):
\begin{equation}
    \label{eq3.5}
    \langle\psi_{\mathbf{w}}|\psi_{\mathbf{x}}\rangle=\text{}_0\bra{1}U^\dagger_{MD}(\{\eta_t\})PS(\boldsymbol{\phi)}PS^\dagger(\boldsymbol{\theta)}U_{MD}(\{\eta_t\})\ket{1}_0.
\end{equation}
We get more insight by explicitly discussing the unitary evolution of the reference state $\ket{1}_0$. In quantum optical computation, we generally use the Heisenberg picture to formulate the temporal evolution of the quantum system, where the input-output relation of the bosonic operators becomes
\begin{equation}
    \label{eq3.6}
    \hat{b}_k=\sum_{j}u_{kj} \hat{a}_j, \text{ and }\quad  \hat{b}^\dagger_k=\sum_{j}u^*_{kj}\hat{a}^\dagger_j ,
\end{equation}
where $(\hat{a},\hat{a}^\dagger)$ and $(\hat{b},\hat{b}^\dagger)$ are the input and the output bosonic operators, respectively, and $u_{kl}$ are the elements of the unitary matrix. To find out the output state after the application of each operator in Eq. (\ref{eq3.5}), we start with the reference state $\ket{1}_1$, a single photon in the first qmode.
\begin{equation}
    \label{eq3.7}
    \ket{\Gamma_1}=\ket{1}_0=\hat{a}^\dagger_0\ket{0}_0.
\end{equation}
The first multiport device provides an output state that has an equal probability of getting a single photon over all qmodes:
\begin{align}
    \ket{\Gamma_2}&=U_{MD}\hat{a}^\dagger_0\ket{0}_0\notag\\
    &=\sum_{k=0}^{N-1}u_{0k}\hat{b}^\dagger_k\ket{0}_k  ,\label{eq3.8}
\end{align}
where we have used the inverse relation of Eq. (\ref{eq3.7}). For the reference state in Eq. (\ref{eq3.7}), we need $U_{MD}$ with all elements of the first column having the value, $u_{k0}=1/\sqrt{N}$ and the output state in Eq.(\ref{eq3.8}) reduces to
\begin{equation}
    \label{eq3.9}
    \ket{\Gamma_2}=\frac{1}{\sqrt{N}}\sum_{k=0}^{N-1}\ket{1}_k.
\end{equation}
Subsequently, the phase shifters are operated as $PS(\boldsymbol{\phi})PS^\dagger(\boldsymbol{\theta})$ to obtain
\begin{align}
    \ket{\Gamma_3}&=PS(\boldsymbol{\phi})PS^\dagger(\boldsymbol{\theta})\ket{\Gamma_2}\notag\\
    &=\frac{1}{\sqrt{N}}\sum_{k=0}^{N-1}e^{i(\theta_k-\phi_k)}\ket{1}_k .\label{eq3.10}
\end{align}
The Hermitian conjugate of the multiport device has to apply finally on the quantum state in Eq. (\ref{eq3.10}):
\begin{equation}
    \ket{\Gamma_4}=U^\dagger_{MD}\ket{\Gamma_3}=\frac{1}{\sqrt{N}}\sum_{k=0}^{N-1}e^{i(\theta_k-\phi_k)}(U^\dagger_{MD}\hat{a}^\dagger_k)\ket{0}_k,
\end{equation}
where the linearity property of the quantum operators is used. Exploiting the inverse relation in Eq. (\ref{eq3.6}), we get
\begin{align}
    \ket{\Gamma_4}&=\frac{1}{\sqrt{N}}\sum_{k=0}^{N-1}e^{i(\theta_k-\phi_k)}\left(\sum_{l=0}^{N-1}u_{lk}\hat{b}_l^\dagger\right)\ket{0}_l\notag\\
    &=\frac{1}{\sqrt{N}}\sum_{l=0}^{N-1}\left(\sum_{k=0}^{N-1}e^{i(\theta_k-\phi_k)}u_{lk}\right)\hat{b}_l^\dagger\ket{0}_l\label{eq3.11}.
\end{align}
When the elements in the first column of $U_{MD}$ have the value $1/\sqrt{N}$, the same will follow for the first row of the Hermitian Conjugate multiport device $U^\dagger_{MD}$, \textit{i.e.}, $u_{0k}=1/\sqrt{N}$ and the above Eq. (\ref{eq3.11}) becomes
\begin{equation}
    \label{eq3.12}
    \ket{\Gamma_4}=\frac{1}{N}\sum_{k=0}^{N-1}e^{i(\theta_k-\phi_k)}\ket{1}_0+\frac{1}{\sqrt{N}}\sum_{l=1}^{N-1}\sum_{k=0}^{N-1}e^{i(\theta_k-\phi_k)}u_{lk}\ket{1}_l.
\end{equation}
Quantum measurement of the single photon on the first spatial qmode returns the identical result as seen in the equations Eqs. (\ref{eq2.5}) and (\ref{eq2.6}):
\begin{align}   \left|\langle\psi_{\mathbf{w}}|\psi_{\mathbf{x}}\rangle\right|^2&=\left|\text{}_0\bra{1}\Gamma_4\rangle\right|^2=\left|\frac{1}{2^n}\sum_{k=0}^{2^n-1}e^{i(\theta_k-\phi_k)}\right|^2  \label{eq3.12a}\\
   &=\frac{1}{2^n}+\frac{1}{2^{2n}}\sum_{j<k}cos\left((\theta_k-\phi_k)-(\theta_j-\phi_j)\right).\label{eq3.13}
\end{align}
This establishes our quantum optical model of an artificial neuron. However, it is important to discuss the architectures of the required optical circuits, namely multiport devices and phase shifters. We start with the implementation of the phase shifters $PS(\boldsymbol{\beta})$, since designing its optical circuit is straightforward. The simplest case is the transformation of a single input bosonic operator $\hat{a}_k$ under the one qmode phase shifter $PS(\beta_k)$:
\begin{equation}
    \label{eq3.14}
    \hat{b}_k=PS^\dagger(\beta_k)\hat{a}_kPS(\beta_k)=e^{-i\beta_k\hat{a}^\dagger_k\hat{a}_k}\hat{a}e^{i\beta_k\hat{a}^\dagger_k\hat{a}_k}=e^{i\beta_k}\hat{a},
\end{equation}
where $e^{i\beta_k\hat{a}^\dagger_k\hat{a}_k}$ is the unitary form of the single mode phase shifter and $\hat{b}_k$ is the output bosonic operator. It easily extends to multi-qmodes phase shifter transformation by applying the single mode phase shifters over all qmodes in  parallel, $PS(\boldsymbol{\beta)=\bigotimes_{k=0}^{N-1}}PS(\beta_k)$, which transforms the multimode input bosonic operator as
\begin{equation}
    \label{eq3.15}
    \begin{pmatrix}
        \hat{b}_0\\\hat{b}_1\\\vdots\\\hat{b}_{N-1}
    \end{pmatrix}=
    \begin{pmatrix}
        e^{i\beta_0}&0&\dots&0\\
        0&e^{i\beta_1}&\dots&0\\
        \vdots&\vdots&\ddots&\vdots\\
        0&0&\dots&e^{i\beta_{N-1}}\\
    \end{pmatrix}
    \begin{pmatrix}
        \hat{a}_0\\\hat{a}_1\\\vdots\\\hat{a}_{N-1}
    \end{pmatrix}.
\end{equation}
The essence is also to propose an optical circuit synthesis algorithm for the implementation of the multiport device (MD). For our work, this multiport creates a superposed state of finding a single photon overall output qmodes, when a single photon is given in the first input spatial qmode. However, we provide a generic algorithm where the probability amplitude of finding a single photon at the output qmodes is different but real.
\begin{figure*}
    \centering
    \includegraphics[width=0.8\linewidth]{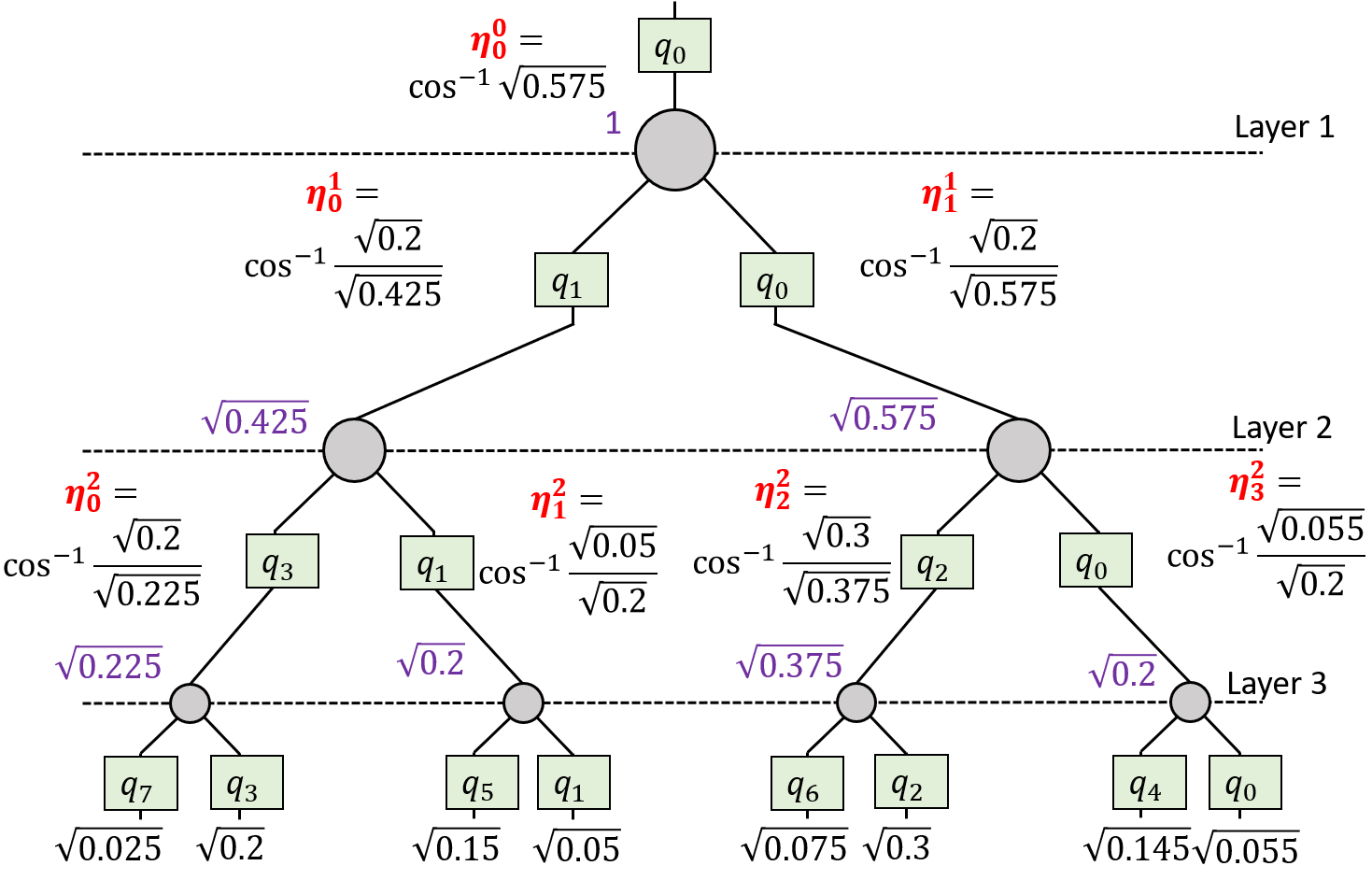}
    \caption{ (Color online) This is the architecture of the quantum optical circuit that generates a real-valued vector $\mathbf{c}$ in Eq. (\ref{eq3.1.2}). Here, dotted lines represents layers. The transmissivity angles $\{\eta_t^l\}$ for the given layer are generated via the algorithm in Sec. (\ref{algorithm1}). The green boxes are the qmodes on which the beam splitters act, whereas the digits in a purple color present components of a temporary vector that are used to calculate the transmissivity angles.}
    \label{fig4}
\end{figure*}

\subsection{Quantum Optical Circuit Synthesis Algorithm}
Consider a \emph{unit-norm} real vector $\mathbf{c}$, whose every component $c_k$ lies between $[0, \infty]$. Our job is to design an MD that gives a quantum state whose probability amplitude encodes the information of $\mathbf{c}$. This multiport device consists of beam splitters (BSs). The generic matrix representation of a BS is
\begin{equation}
    \label{eq3.1.1}
    \begin{pmatrix}
        \hat{b}_1\\ \hat{b}_2
    \end{pmatrix}=\begin{pmatrix}
        \cos \eta&-e^{-i\xi}\sin\eta\\
        e^{i\xi}\sin\eta&\cos\eta
    \end{pmatrix}\begin{pmatrix}
        \hat{a}_1 \\ \hat{a}_2
    \end{pmatrix},
\end{equation}
where $\cos\eta\text{ and }e^{-i\xi}\sin\eta$ are the transmissivity and reflectivity amplitudes of the BS, respectively. For encoding a real-valued vector $\xi=0$. Now, we need to get an algorithm that returns transmissivity angles $\{\eta_t\}$ for the MD when a vector $\mathbf{c}$ is provided.\\
\begin{algorithm}
\caption{An algorithm for finding the transmissivity angles of the beam splitters}
\begin{algorithmic}[1]
\label{algorithm1}
\renewcommand{\algorithmicrequire}{\textbf{Input:}}
\renewcommand{\algorithmicensure}{\textbf{Output:}}
\REQUIRE A vector $\mathbf{c}$ with dimension $N$
\ENSURE Returns the transmissivity angles for beam splitters of MD as a nested list
\STATE $L=\lceil log_2N\rceil$  \# return the number of layers of MD
\STATE vector=[$\mathbf{c}$]  \# pass the actual vector $\mathbf{c}$ to a new variable vector as a list object
\STATE transmission angles=[\text{ }] \# creates an empty list object variable, where transmission angles are stored
\FOR{$l\leftarrow 0$ to $L$}
    \STATE temporary vector=[\text{ }]\# creates an empty list object variable, where each layer's vector stored
    \STATE temporary transmission angles=[\text{ }]\# creates an empty list object variable, where each layer's transmission angles stored
    \IF{$\text{length}(N)\%2==0$}
        \STATE pass
    \ELSE
        \STATE vector[$l$]=vector[$l$]+[0] \# add a zero element in the end of the list
    \ENDIF
     \FOR{$t\leftarrow 0$ to $\text{length}(\text{vector}[l])/2$}
            \STATE temporary vector component=$\sqrt{\text{(vector}[l][2t])^2+\text{(vector}[l][2t+1])^2}$
            \STATE temporary vector appends temporary vector component
            \STATE temporary transmission angles elements= $\cos^{-1}\left(\frac{\text{vector}[l][2t+1]}{ \text{temporary vector component}}\right)$
            \STATE temporary transmission angles appends temporary transmission angles elements
        \ENDFOR
        \STATE vector appends temporary vector
        \STATE transmission angles appends temporary transmission angles
\ENDFOR
\end{algorithmic}
\end{algorithm}
\begin{figure*}
    \centering
    \includegraphics[width=0.8\linewidth]{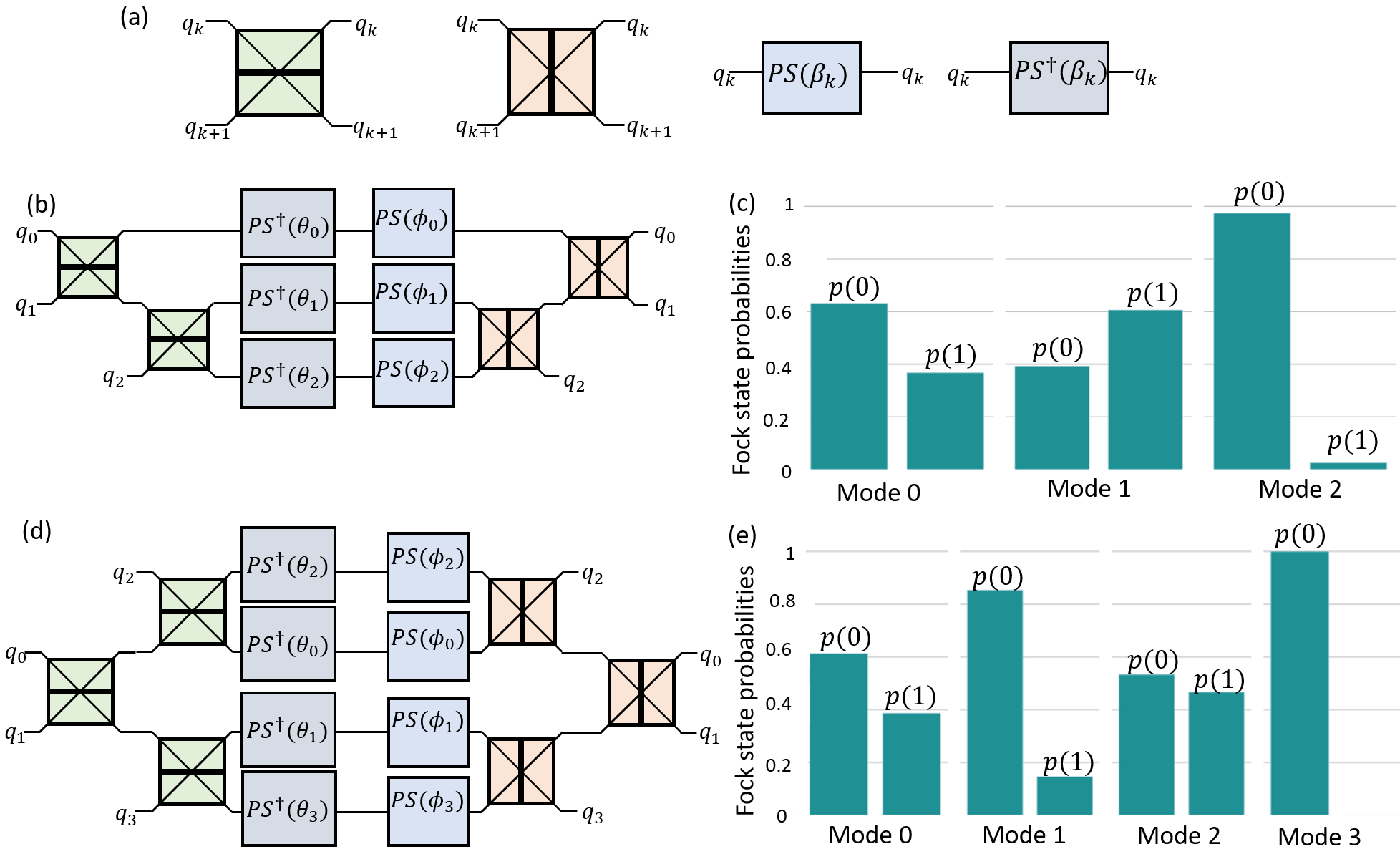}
    \caption{Linear optical elements are shown in the schematic (a) in the given order, such as two-qmodes beam splitter and its conjugate version; a single mode phase shifter and its conjugate version. Figures (b) and (d) are quantum optical circuits of three- and four-dimensional input data quantum neurons, respectively. The results of the numerical simulation of the linear optical circuits are shown in the histogram plots (c) and (e), respectively.}
    \label{fig5}
\end{figure*}

Two important points need to be highlighted regarding this algorithm: first, sometimes we may get the transmissivity angles equal to $\pi/2$ and we need not to add beam splitters to the MD. Second, the lower layer beam splitters' transmissivity angles $\{\eta^l_t\}$ are found from the algorithm (\ref{algorithm1}) and then, we obtain for other successive layers.

The obtained transmissivity angles from this algorithm will help us to find the corresponding quantum optical structure. The quantum optical structure looks similar to a pyramid structure, where $L$ layers stack on each other. We take an example by considering the following unit vector
\begin{equation}
 \label{eq3.1.2}
    \mathbf{c}=(\sqrt{0.025}, \sqrt{0.2}, \sqrt{0.15}, \sqrt{0.05}, \sqrt{0.075},\sqrt{0.3}, \sqrt{0.145}, \sqrt{0.055}) ,   
\end{equation}

whose information is encoded in the quantum optical state. Then, this vector is fed into the algorithm to get the transmission angles for each layer. The prototypical quantum optical circuit is given in Fig. \ref{fig4}, corresponding to the obtained transmissivity angles. Let's compute the transformation matrix corresponding to the quantum optical circuit in Fig. \ref{fig4}. An equivalent transmission matrix for the first layer is required before moving towards the successive layers. The mathematical formalism to simulate the unitary matrix with the linear optical network can be used \cite{reck1994experimental}. However, this formalism has to be used in reverse order, \textit{i.e.}, preparing a transformation matrix from the given structure of a linear optical network. Once the transformation matrix is computed for each layer, their matrix product is evaluated in proper sequence to obtain the resultant transformation matrix for this quantum optical circuit as\vspace{1 cm}
\begin{align*}
    &BS(q_0, q_4)BS(q_1, q_5)BS(q_2, q_6)BS(q_3, q_7)\times BS(q_0, q_2)BS(q_1, q_3)\times BS(q_0, q_1)=\\
    & \hspace{0.8 cm} \begin{pmatrix}
        \sqrt{0.055} & -\sqrt{\frac{0.0234}{0.575}} & -\sqrt{\frac{0.021}{0.115}} & 0 & -\sqrt{\frac{0.145}{0.2}} & 0 & 0 & 0\\
        \sqrt{0.05} & \sqrt{\frac{0.029}{0.445}} & 0 & \sqrt{\frac{0.011}{0.089}} & 0 & -\sqrt{\frac{0.15}{0.2}} & 0 & 0\\
        \sqrt{0.3} & -\sqrt{\frac{0.127}{0.575}} & \sqrt{\frac{0.06}{0.129}} & 0 & 0 & 0 & -\sqrt{\frac{0.075}{0.375}} & 0\\
        \sqrt{0.2} & \sqrt{\frac{0.035}{0.445}} & 0 & \sqrt{\frac{0.04}{0.1}} & 0 & 0 & 0 & -\sqrt{\frac{0.025}{0.215}}\\
        \sqrt{0.145} & -\sqrt{\frac{0.062}{0.575}} & -\sqrt{\frac{0.054}{0.115}} & 0 & \sqrt{\frac{0.055}{0.2}} & 0 & 0 & 0\\
        \sqrt{0.15} & \sqrt{\frac{0.086}{0.445}} & 0 & -\sqrt{\frac{0.034}{0.089}} & 0 & \sqrt{\frac{0.05}{0.2}} & 0 & 0\\
        \sqrt{0.075} & -\sqrt{\frac{0.032}{0.575}} & \sqrt{\frac{0.015}{0.216}} & 0 & 0 & 0 & \sqrt{\frac{0.3}{0.375}} & 0\\
        \sqrt{0.025} & -\sqrt{\frac{0.004}{0.445}} & 0 & \sqrt{\frac{0.005}{0.107}} & 0 & 0 & 0 & \sqrt{\frac{0.2}{0.275}}
    \end{pmatrix}
\end{align*}
where parentheses associated with each BS tell about those two qmodes on which it is acting. In the architecture in Fig. \ref{fig4}, the qmodes (shown within the box in Fig. \ref{fig4}) carry the information of the components of $\mathbf{c}$ in Eq. (\ref{eq3.1.2}), but of different order.

\subsection{Numerical Simulation of Quantum Optical Neuron and Circuit Cost}

We further validate the quantum optical circuit synthesis algorithm for the proposed quantum optical model of an artificial neuron by numerical simulation using Strawberry Fields (a Python-based photonic simulation kit) \cite{killoran2019strawberry}. As mentioned earlier, we need only two types of linear passive optical elements to build our required circuit, such as beam splitter and phase shifters, as shown in Fig. \ref{fig5} (a). Here, their conjugate version is displayed to present an easier schematic of the quantum optical circuit. We prepare two optical circuits corresponding to three- and four-dimensional input data as shown in Fig. \ref{fig5} (b) and Fig. \ref{fig5} (d), respectively. The three-dimensional input and weight are respectively given by $\mathbf{x}=[\pi/7, \pi/3, \pi/2]$ and $\mathbf{w}=[\pi/2, \pi/8, \pi/6]$, whereas for four-dimensional input data, the input is $\mathbf{x}=[\pi/6, \pi/3, \pi/2, \pi/5]$ and weight is $\mathbf{w}=[\pi/2, \pi/8, 0, 0]$. The three-dimensional input data is chosen in the quantum optical model to avoid the padding with one extra zero to make the dimension 4. Once we run both the optical circuits on the Strawberry Fields photonic simulator, we obtain the output as the Fock state probabilities of getting no photon $ p(0)$ and a single photon $p(1)$ of each mode which is shown in Fig. \ref{fig5} (c) and Fig. \ref{fig5} (e). The probability of getting a single photon $p(1)$ in mode 0 is the output of the quantum neuron. The resultant probability $p(1)$ in mode 0 for three- and four-dimensional input data are $0.368$ and $0.386$, respectively, which conform the outcome of our calculation explicitly  from the formula Eq. (\ref{eq2.6}). Hence, our quantum optical model as well as the corresponding circuit synthesis algorithm get validated.

It is also crucial for practical implementation to find out the comparative merit of our quantum optical neuron model against the qubit-based model through their cost metrics. The circuit costs are calculated in terms of the dimension of the input data $N$.
\begin{enumerate}
     \item[(a)] \textbf{Circuit size}: In our case, the circuit size of a quantum optical neuron is comparable to that of a qubit-based circuit, which is $(2^{\lceil log_2 N\rceil}+N-2)$. This number may further increase by $N$ if the input and the weight involve two layers of phase shifters. We have avoided it by merging the phase shifters during the implementation of diagonal operators.
     \item[(b)] \textbf{Circuit depth}: The depth of the optical circuit is $2(\lceil log_2 N\rceil+1)$, which is exponentially smaller than the circuits, driven by qubit-based circuit synthesis algorithms.
     \item[(c)] \textbf{Circuit width}: The width of the optical circuit is $log_2N$, which is always one less than a qubit-based circuit.
 \end{enumerate}
Hence, the above discussion on the circuit costs makes our validated quantum optical model of an artificial neuron practically worthy. Moreover, the presented linear quantum optical circuits preclude expensive resource like $Cnot$ gates.

\section{Conclusions}
 \label{sec4}
We have proposed a quantum optical implementation of the quantum neuron, building upon the qubit-based architecture introduced by Mangini \textit{et al.} \cite{mangini2020quantum}. We began by analyzing two quantum circuit synthesis algorithms, capable of realizing quantum neurons with arbitrary dimensionality. We have also developed a corresponding quantum optical architecture that faithfully replicates the qubit-based model. To ensure correctness and functionality, the quantum optical model and the associated synthesis algorithm are validated. A comprehensive circuit cost analysis demonstrates that the quantum optical neuron offers a reduction in resource requirements compared to its qubit-based counterpart, emphasizing its practical viability. Furthermore, the model is capable of implementing both phase-encoded and real-valued quantum neurons \cite{monteiro2021quantum}, thereby establishing it as a flexible and general-purpose framework for quantum neural computation.
 

\section{Acknowledgement}
UR acknowledges the support from SERB-CRG, GoI and also DST under the National Quantum Mission (Quantum Computing).




\end{document}